\documentclass[12pt,aps,prb,floats,eqsecnum,psfig]{revtex4}
\def\lapprox{\,\raise0.4ex\hbox{$<$}\kern-0.8em\lower0.7ex\hbox{$\sim$}\,}
\def\gapprox{\,\raise0.4ex\hbox{$>$}\kern-0.8em\lower0.7ex\hbox{$\sim$}\,}

\begin{document}
\title{Relaxation of Inter-Landau-level excitations\\
in the Quatised Hall Regime}

\medskip
\author{\underline{S.
Dickmann${}^1$}${}\!$\renewcommand{\thefootnote}{\fnsymbol{footnote}}
\footnotemark[1]
 and Y. Levinson${}^2$}
\address{
${}^1$Institute for Solid State Physics of Russian Academy of
Sciences,\\
142432 Chernogolovka, Moscow District, Russia}
\address{${}^2$Department of Condensed Matter Physics, The Weizmann
Institute of Science,\\
76100 Rehovot, Israel\\
\parbox{14.2cm}
{\rm
\bigskip
\noindent
Relaxation of collective plasmon inter-Landau-level excitation
is determined by emission of LO-phonons or by Auger-like processes
when this emission is suppressed off the magneto-phonon resonance
conditions. The decay of ``one-cyclotron" magneto-plasmons with
wave-vectors near the roton minimum is studied under the condition
of filling $\nu=1$. Some features of this relaxation should be
helpful for the experimental detection of the magnetorotons in a
strongly correlated 2D electron gas (2DEG).
\vspace{-2mm}
\begin{flushleft}
\end{flushleft}}}
\maketitle
\renewcommand{\thefootnote}{\fnsymbol{footnote}}
\footnotetext[1]{Fax: +7 096 576 4111; e-mail: dickmann@issp.ac.ru}

\bigskip
In recent two decades considerable  interest has been focused on the
collective excitations in a strongly correlated 2DEG under the
quantum Hall regime conditions. The study of such excitation provide
a way to determine the fundamental properties of a 2DEG which eventually
explain its relaxation and transport features. For integer filling
$\nu$ the calculation of the exciton-like spectra (which are in fact of
Bose type) in the limit of high
magnetic fields reduces to an exactly solvable problem [1-3].
At the same time the direct experimental discovery of such excitons
(spin-flip waves, magneto-plasmons (MPs) without and with spin
flip) presents certain difficulties. The inter-Landau-level MPs
were observed in the works [4-5] by means of inelastic
light-scattering. However, a massive breakdown in wave-vector
conservation implied for this detection is not understood as
yet.

The presented paper is devoted to the properties of MP relaxation
(MPR) which might help indirectly to reveal the presence of MP
excitations in 2DEG, namely: the two types of MPR for the filling
$\nu=1$, which are studied below, should give rise to the special
features in the hot luminescence and Raman scattering signals from
2DEG. In particular, a nonmonotonic dependence on the magnetic field
has to be expected for the intensity of the hot luminescence which
arises due to electron relaxation from the first and the second
Landau levels (LLs).

We concern only the MPs without spin flip,
i.e. with the energies
$$
  \epsilon_{ab}(q)=\hbar\omega_c(n_b-n_a)+ {\cal E}_{ab}(q)
  \qquad (n_b>n_a),                                   \eqno(1)
$$
where $\omega_c$ is the cyclotron frequency, $n_a$ and $n_b$ are the
numbers of initially (in the ground state) occupied and unoccupied
LLs respectively. Having a Coulomb origin the energy
${\cal E}_{ab}$ is of the order of or smaller than
$E_C=e^2/\kappa_0 l_B$ which is a characteristic energy of
electron-electron interaction in 2DEG ($l_B$ is the magnetic
length, $\kappa_0$ is the dielectric constant). We should
especially pay
attention for MPs in such portions of their spectrum where the
density of states becomes infinite (i.e. where
$d\epsilon_{ab}/dq=0$) or/and the wave-vector $q$ is equal to
zero. When so doing we take into account the definitive role
of an experimental test. Indeed, the features in light-scattering
spectra
[4-5] attributed to the one-level excitations (where $n_b-n_a=1$)
are only detected  in the vicinity of $q=0$ (therewith
${\cal E}_{ab}=0$ but $d{\cal E}_{ab}/dq\not=0$) or near their
roton minimum. In the latter case the interaction energy is
$$
  {\cal E}_{01}\approx \varepsilon_{0} + (q-q_0)^2/2M,
  \quad     |q-q_0|\ll q_0                           \eqno(2)
$$
(the index $ab$ is specified by replacing it with $n_an_b$, since the
spin state does not change in our consideration). We consider the
case where
$n_a=0$ and $n_b=1$ with filling $\nu=1$; then in this equation
$q_0\approx 1.92/l_B$, $M^{-1}\approx 0.28\,E_{C}l_{B}^2$ and
$\varepsilon_0 \approx 0.15\,E_{C}$ in the strict 2D limit (namely
if 2DEG thickness $d$ satisfies the condition $d\ll l_B$). Actually
the MP spectra depend on $d$ but their shape do not change
qualitatively.

\smallskip
{\bf A}. First we consider the magnetophonon resonance conditions
when the energy $\epsilon_{01}$ defined by Eq. (1) is equal to
the LO-phonon energy $\hbar\omega_{LO}=35\,$meV [6].
The resonance with $q=0$ when $\omega_c=\omega_{LO}$ is just a
consequence of the Kohn theorem and does not demonstrate the
presence of MP excitations in the system.
Therefore, we consider the case of magnetoroton relaxation
when MP energy is transferred to the emitted optic phonon.
Then in the vicinity of $q\approx q_0$ we should expect the resonance
if $\hbar(\omega_{LO}-\omega_c)\approx {\cal
E}_{01}(q_0)=\varepsilon_0$. In the strict 2D limit this condition
leads to the resonant magnetic field $B=19\,$T instead of $21\,$T
corresponding to the case of $\omega_c=\omega_{LO}$. Note that in
this case the MPR has to be accelerated. As a result, if the hot
luminescence from the 1-st LL is measured, then a fall in its
intensity in the resonant magnetic field should be detected.

We calculate now the rate of this relaxation. Describing the
states of the system in terms of the Excitonic Representation which
means the transition from electron annihilation and creation
operators to the excitonic ones $Q_{ab{\bf q}}$ and $Q_{ab{\bf q}}^+$
(see Refs. [7-10]), we should therefore study the transition
between electron states $|i\rangle=Q_{ab{\bf q}}^+|0\rangle$ and
$\langle f|=\langle 0|$ ($\langle 0|$ is the ground state). The
relevant matrix element calculation
$$
  {\cal M}_{ab{\bf q}}=
  \langle 0|{\cal H}_{e,ph}|Q_{ab{\bf q}}^+|0\rangle
                                             \eqno(3)
$$
assumes the Excitonic Representation for Hamiltonian of
electron-phonon interaction,
$$
{\cal H}_{e,ph}=\displaystyle{\frac{1}{L}\left(\frac{\hbar}{L_z}
\right)^{1/2}}\,
\sum_{{\bf k}}
{\tilde U}_{opt}^*({\bf k})
H_{e,ph}({\bf q})\;,           \eqno(4)
$$
where $L\times L\times L_z$ are the sample sizes, ${\bf k}=({\bf
q}, k_z)$ is the phonon wave-vector, ${\tilde U}_{opt}({\bf k})=
\gamma(k_z){U}_{s}({\bf k})$ is the renormalized
Fr\"ohlich vertex ($\gamma$ is the size-quantised form-factor
[8]), namely: $|U_{opt}|^2=2\pi
e^2\omega_{LO}/\overline{\kappa}k^2$ (the standard
notation for the reduced dielectric constant ${\overline\kappa}^{-1}
= \kappa^{-1}_{\infty}-\kappa^{-1}_0$ is used). The relevant
representation for $H_{e,ph}$ operating on the electron
states can be obtained similar to the case of
exciton-acoustic-phonon interaction [8]:
$$
  H_{e,ph}=\frac{L}{l_B\sqrt{2\pi}}
  \left[h_{n_an_b}({\bf q})Q_{ab{\bf q}}
  + h^*_{n_an_b}(-{\bf q})Q^+_{ab\,-\!{\bf q}}\right].\eqno(5)
$$
Here $n_b\geq n_a$, and
$$
  h_{n_an_b}({\bf q})=(n_a!/n_b!)^{1/2}(q_+l_B)^{n_b-n_a}
  e^{-q^2l_B^2/4}L_{n_a}^{n_b\!-\!n_a}(q^2l_B^2/2)\,,       \eqno(6)
$$
where $L_n^j$ is Laguerre polynomial, $q_{\pm}={\mp
i}{2^{-1/2}}(q_x\pm iq_y)$. Now exploiting the relevant commutaion
rules for the excitonic operators (see Refs. [8] and [10]) we can find the
matrix element (3) appropriate in our case,
$$
  {\cal M}_{01{\bf q}}=(\hbar/2\pi\!L_z)^{1/2}
  U_{opt}^*({\bf k})h_{01}({\bf q})/l_B,
                                                          \eqno(7)
$$
and then the MPR rate
$$
  R_{ph}=\sum_{{\bf q}',k_z}\frac{2\pi}{\hbar}|{\cal M}_{01{\bf q}'}|^2
  \delta(\epsilon_{01}-\hbar\omega_{LO})=\frac{e^2L^2\omega_{LO}}
  {4\overline{\kappa}|d{\cal E}_{01}/dq|}q^2e^{-q^2l_B^2/2}
  \overline{n}(q).
                                                          \eqno(8)
$$
In the last expression $q$ is the root of equation
$\epsilon_{01}(q)=\hbar\omega_{LO}$, and $\overline{n}(\mbox{\boldmath
$q$})$
is the occupation number of 01MPs.

Formally the result (8) becomes
infinite when $q=q_0$. However, the real magnitude of $R_{ph}$ in the
vicinity of $q_0$ can be found from the analysis of the
homogeneity breakdown due random impurity potential $U({\bf r})$.
Assuming $U({\bf r})$ to be smooth (its
correlation length $\Lambda\gg l_B$) one can find that the energy
correction for any $ab$MP is in the dipole approximation
$\delta {\cal E}=-\hbar {\bf q}{\bf v}_d$,
where ${\bf v}_d=({\hat z}\times\nabla U({\bf r}))
l_B^2/\hbar$ is the drift velocity (see Refs. [3] and [10]).
This energy is an inhomogeneous broadening of the MP
energy and has to be added to ${\cal E}_{ab}$. The random potential
correction plays no significant role if
$|d{\cal E}_{ab}/dq|\gg l_B^2\left|{\bf \nabla}U\right|$,
which means that the electron-hole Coulomb
interaction is stronger than the force the electron and the hole
are subjected to in the random potential.
In other words, the derivative $|d{\cal E}_{ab}/dq|$ in Eq. (8) is
limited from below by  $l_B^2\left|{\bf
\nabla}U\right|\sim l_B^2\Delta/\Lambda$, where $\Delta$ is the
random potential amplitude. Thus we obtain the rate near $q_0$
per unit area:
$$
  \left[R_{ph}/L^2\right]_{\mbox{max}}\sim\frac{e^2\Lambda\omega_{LO}}
  {4\overline{\kappa}l_B^2\Delta}q_0^2e^{-q_0^2l_B^2/2}
  \overline{n}(q_0).                          \eqno(9)
$$
The roton minima broadening due to inhomogeneity is $|{\bf q}-{\bf q}_0|\sim
(2M\delta{\cal E})^{1/2}$. Estimating magnetoroton density as $N\simeq
\overline{n}q_0(2M\delta{\cal E})^{1/2}$ and setting $dN/dt$
equal to decay rate (9) we find the characteristic relaxation time
$\tau_{ph}=\overline{n}dt/d\overline{n}$ which turns out to be of the order
of
$$
  \tau_{ph} \sim 4\exp{(q_0^2l_B^2/2)}
  \left(\frac{\Delta}{\Lambda}\right)^{3/2}
  \left(
  \frac{2M}{q_0}\right)^{1/2}\frac{\overline{\kappa}l_B^3}
  {e^2\omega_{LO}} \sim 0.1\div 0.01\; \mbox{ps}. \eqno(10)
$$
(we assume that $B=10\,$T, $\Delta\simeq 1\,$meV, $\Lambda\simeq 50\,$nm).

\smallskip
{\bf B}. Naturally, the above results should be compared with
the analogous ones in the case when emission of LO-phonons is
suppressed off the resonance conditions. Generally, the
MPR mechanism seems to be determined by many-phonon emission.
However,
a certain additional relaxation channel exists precisely for the
considered magnetorotons. A coalescence of two of them with their
conversion into a single MP of the ``two-cyclotron" plasmon mode (with
$n_a=0$, $n_b=2$) turns out to be energetically allowed because
``by chance" the difference $\delta={\cal
E}_{02}(0)-2\varepsilon_0$ is numerically very small. Namely, in
the strict 2D limit $\delta \approx 0.019\,E_{C}\simeq 3\div 4\,$
K for $B=10\div 20\,$T. This coalescence leads to an Auger-like
MPR process because as a result the total number of excited
electrons decreases as well as the total number of MP excitations.
The dependence ${\cal E}_{02}(q)$ is
nonmonotonous, but in the range $0<ql_B<2.5$ does not change by more
than $0.07\,E_c$. Nevertheless, it would be preferable observe the
generated ``two-cyclotron" MP in
the state with small 2D wave-vector, because in this case the
generated MP could be detected by anti-Stokes Raman scattering
similar to the experiments of Refs. [4-5].

We calculate the decay rate due to such an Auger-like process,
$$
 {\cal R}={1\over 2}\sum_{{\bf q}_1,{\bf q}_2}\frac{2\pi}{\hbar}
  \left|{\cal M}({\bf q}_1,{\bf q}_2)\right|^2
  \overline{n}({\bf q}_1) \overline{n}({\bf q}_2)
 \delta\left[{\cal E}_{01}({\bf q}_1)+{\cal E}_{01}({\bf q}_2)
  -{\cal E}_{02}({\bf q}_1+{\bf q}_2)\right]\,,       \eqno(11)
$$
where the required matrix element of the considered conversion is
$$
  {\cal M}({\bf q}_{1},{\bf q}_{2})
  =\left\langle 0\left|Q_{02\,{\bf q}_1\!+\!{\bf q}_2}\left|
  H_{int}\right|Q^+_{01{\bf q}_{1}}Q^+_{01{\bf q}_{2}}\right|0\right\rangle.
                                                       \eqno(12)
$$
Here bra and ket states are orthogonal, and $H_{int}$ is the Coulomb
interaction Hamiltonian of 2DEG.
Rewriting $H_{int}$ in the Excitonic Representation we should take
into account that within the framework of the exploited high magnetic
field approximation it is sufficient to keep in $H_{int}$ only the
terms which commute with the Hamiltonian of noninteracting
electrons and therefore conserve cyclotron part of the 2DEG energy.
When so doing we find that the only term which gives the contribution
to the matrix element (12) is
$$
  H_{int}'=\frac{1}{2\pi l_B^{2}}\sum_{\bf q}V(q)
  h_{10}({\bf q})h_{21}^*({\bf q})Q^+_{12{\bf q}}Q_{01{\bf q}},
                                                     \eqno(13)
$$
where $V(q)$ is the 2D Fourier component of the Coulomb potential
averaged with the wave function in the ${\hat z}$ direction
(so that in the strict 2D limit: $V(q)=2\pi l_BE_C/q$).
Substituting the operator (13) for $H_{int}$ into Eq. (12) and
employing the special commutation rules for the excitonic
operators ( see Refs. [8] and [10]) one can calculate the matrix
element (12) for arbitrary ${\bf q}_{1}$ and ${\bf q}_{2}$. We
will later need this quantity only when
${\bf q}_{1}\approx -{\bf q}_{2}$ and
$q_1\approx q_2\approx q_0$. In this case
${\cal M}({\bf q}_{0},-{\bf q}_{0})=-2(2\pi)^{1/2}\mu l_B/L$, where
in the strict 2D limit $\mu\approx 0.062E_C$.

To calculate the depopulation rate of 01MPs (11) one has additionally
to know the $\overline{n}({\bf q}_1)$ distribution and the
appropriate phase area $A$ for the relevant final wave-vectors of
02MPs ${\bf q}={\bf q}_1+{\bf q}_2$. When this area is
sufficiently small, namely if $\pi q^2\leq A\ll 4\pi^3M\delta$,
the result is
$$
  {{\cal R}}/{L^{2}}=
  \displaystyle\frac{\overline{n}^2(q_0)\mu^2l_B^2q_0}{2\pi\hbar}
  \left({M\over\delta}\right)^{1/2}A\,.  \eqno(14)
$$
This is just the case for the rate (11) when 02MP
creation occurs in the phase area relevant for anti-Stokes inelastic
backscattering. Then the role of the random potential in determining
the value of $A$ is crucial. Indeed, if $ql_{B}<1$,
one can get an estimate
$d{\cal E}_{02}/dq\sim E_Cq^2l_B^3$ (see Ref. [2]), and
the uncertainty in $q$ due to disorder turns out to be
${\tilde q}\sim (\Delta/E_{c})^{1/2}(\Lambda l_{B})^{-1/2}$
(for the adopted numerical parameters we find ${\tilde q}\sim
10^5\,$cm${}^{-1}$). The quantity $\pi {\tilde q}^2$ should be
substituted into Eq. (14) for $A$.

If we wish to obtain the total rate of the Auger-like MP
relaxation, then with the help of
Eq. (11) a more complicated summation has to be
fulfilled. Nevertheless, in this case Eq. (15) may be also
employed for the approximate estimation if we substitute there
$A\sim \pi l_B^{-2}$. Then estimating the 01MP density near their roton
minima as $N\simeq
\overline{n}(q_0)q_0(2M\delta{\cal E})^{1/2}$ and setting $dN/dt$
equal to the relevant total rate of the coalescing 01MPs we find
$$
  \tau_{Aug}=\overline{n}dt/d\overline{n}\sim
  \frac{2\hbar}{\overline{n}\mu^2}
  \left(\frac{2\delta\Delta l_B}{\Lambda}\right)^{1/2} \sim
  1/\overline{n}\,\mbox{ps}\,.                              \eqno(15)
$$
This time is by about a factor of 100 is longer than that given by Eq.
(10). On the other hand, the considered Auger-like process
is certainly the dominant relaxation channel in the case of
01-magnetorotons, if the magneto-phonon resonant conditions are not
met.

It is very important that the studied type of MP relaxation can
reveal an additional possibility for the indirect experimental
detection of the magnetorotons. Indeed, if one somehow excites
01MPs near their roton minima, then one could simultaneously observe
the 02MPs (and therefore electrons at the 2-nd LL). It
seems such an observation might be performed by means of
anti-Stokes Raman scattering or by means of hot luminescence from
the 2-nd LL. Note also that if the 1-st
LL turns out in the vicinity of the LO-phonon energy, one
might observe the decrease of the hot luminescence signal from the
2-nd LL in the field $B$ corresponding to the
magnetoroton-phonon resonance studied above. This correlation
between the 1-st LL excitations and the
2-nd LL hot luminescence would be an evidence
of the Auger-like process and therefore of the
magnetoroton existence.

Finally, note that the deviation of the filling from 1 should
nevertheless qualitatively retain the same picture of the considered
MPR as long as this deviation does not reach the point
where the 2DEG excitation
spectrum is drastically renormalized (i.e. the fractional quantum
Hall effect conditions arise).

The work is supported by the MINERVA Foundation and by the Russian
Fund for Basic Research.

\medskip
\noindent
{\bf References}

\noindent
{} 1.
{Yu. A. Bychkov, S. V. Iordanskii, and G. M. \'Eliashberg},
JETP Lett. {\bf 33}, 143  (1981).

\noindent
{} 2.
{C. Kallin, and B. I. Halperin}, Phys. Rev. B {\bf 30}, 5655 (1984).

\noindent
{} 3.
{C. Kallin, and B. I. Halperin}, Phys. Rev. B {\bf 31}, 3635 (1985).

\noindent
{} 4.
{A. Pinczuk, {\it et al.}},
Phys. Rev. Lett. {\bf 68},  3623  (1992); ibid. {\bf 70},  3983  (1993).

\noindent
{} 5.
{A. Pinczuk, {\it et al.}},
Semicond. Sci. Technol. {\bf 9},  1865  (1994).

\noindent
{} 6.
We restrict ourselves to the bulk optic-phonon
modes in GaAs. In this case only the
longitudinal phonons interact with
of the conduction band electrons.

\noindent
{} 7.
{A. B. Dzyubenko, and Yu. E. Lozovik}, Sov. Phys. Solid State {\bf 26},
938 (1984).

\noindent
{} 8. {S. Dickmann, and S. V. Iordanskii}, JETP {\bf 83}, 128 (1996).

\noindent
{} 9. {S. Dickmann}, Physica B {\bf 263-264}, 202 (1999).

\noindent
10. {S. Dickmann, and Y. Levinson}, Phys. Rev. B {\bf ??}, (1996).

\end{document}